# Reaction-Diffusion Degradation Model for Delayed Erosion of Cross-Linked Polyanhydride Biomaterials

Sergii Domanskyi,[†] Katie L. Poetz,[‡] Devon A. Shipp,[‡,§,*] Vladimir Privman[†,§,*]

[†]Department of Physics, [‡]Department of Chemistry and Biomolecular Science, [§]Center for Advanced Materials Processing, Clarkson University, Potsdam, New York 13699

**Absrtact:** We develop a theoretical model to explain the long induction interval of water intake that precedes the onset of erosion due to degradation caused by hydrolysis in the recently synthesized and studied cross-linked polyanhydrides. Various kinetic mechanisms are incorporated in the model in an attempt to explain the experimental data for the mass loss profile. Our key finding is that the observed long induction interval is attributable to the nonlinear dependence of the degradation rate constants on the local water concentration, which essentially amounts to the breakdown of the standard rate-equation approach, potential causes for which are then discussed. Our theoretical results offer physical insights into which microscopic studies will be required to supplement the presently available macroscopic mass-loss data in order to fully understand the origin of the observed behavior.

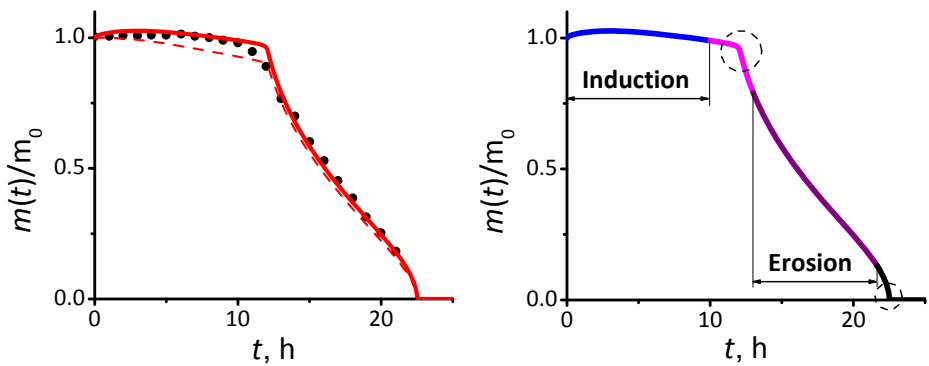

**Graphical Abstract:** Delayed erosion of highly cross-linked polyanhydrides is attributable to the nonlinear dependence of the degradation rates on water concentration.

**Keywords:** crosslinked polyanhydrides, erosion, kinetic model, reaction-diffusion

---

[*]Corresponding authors' e-mails: dshipp@clarkson.edu, privman@clarkson.edu

# PUBLICATION INFORMATION

Journal Issue Cover: Results of this work have been highlighted by the Editors of Phys. Chem. Chem. Phys., by using a compilation of images for a *cover of Issue 20* of Volume 17:

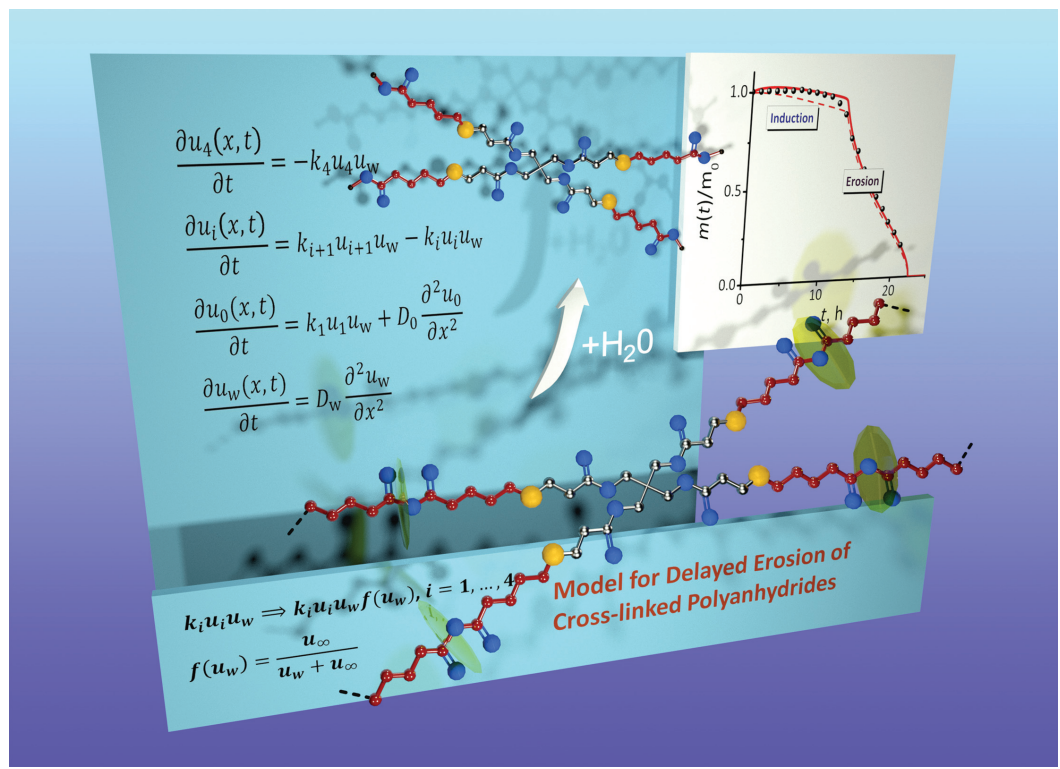

**Showcasing research by V. Privman, D. A. Shipp and colleagues at Clarkson University, Potsdam, New York, USA**

Title: Reaction-diffusion degradation model for delayed erosion of cross-linked polyanhydride biomaterials

This work reports a theoretical model for the long induction interval of water intake that precedes the onset of erosion due to degradation caused by hydrolysis in the recently synthesized cross-linked polyanhydrides. Various kinetic mechanisms are incorporated in the model in order to explain the experimental data for the mass loss profile. The key finding is that the observed long induction interval is attributable to the nonlinear dependence of the degradation rate constants on the local water concentration, which essentially amounts to the breakdown of the standard rate-equation approach.

**As featured in:**

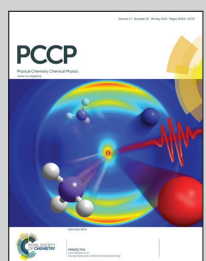

See Devon A. Shipp, Vladimir Privman *et al.*, *Phys. Chem. Chem. Phys.*, 2015, **17**, 13215.

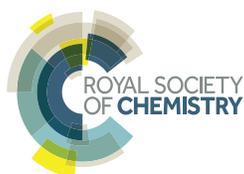

www.rsc.org/pccp

Registered charity number: 207890

# 1. INTRODUCTION

Biodegradable polymers, specifically, polyanhydrides, with tailored chemical and physical properties find diverse applications largely because of their ability to undergo degradation and erosion.[1-5] The use of modeling allows for a better understanding of the degradation and erosion mechanisms of different polymeric biomaterials.[6-16] Here we report a modeling approach aimed at explaining a recently experimentally found[17,18] long induction time of water intake preceding the onset of erosion in cross-linked polyanhydrides synthesized via thiol-ene polymerization. This property is important because it enables additional control in applications, such as drug release.[19,20] The experimental data used to test our model were reported in Ref. 17 that provides the details of the experiments, specifically, the macroscopic mass-loss profile. In our model the observed long delay time in the onset of mass loss is attributed to an interesting kinetic effect of the breakdown of the rate-equation description usually assumed for diffusion-controlled reactions. This allows us to offer physical insights into which microscopic morphological properties should be probed in future experiments.

Predictable degradation and biocompatibility of degradable polymeric biomaterials are required for medical applications.[5,21-24] The former is crucial for drug delivery capsules. The latter is important for orthopedic applications, and some polyanhydrides have compressive strengths similar to the human cortical bone.[24,25] Degradable polymer implants potentially offer benefits such as avoidance of multiple surgeries, elimination of stress shielding, absence of corrosion, and incorporation of curing drugs[22] for slow delivery as the implant degrades and erodes. Other uses could include tissue engineering[5] and bio-adhesives.[23]

Recent experiments on cross-linked polyanhydrides[17] prepared by thiol-ene photopolymerization studied several aspects of the degradation kinetics of such polymers in an aqueous environment. The experimentally measured quantities included the mass-loss profile and release rate of a model drug (a hydrophilic dye), and rate of hydrolysis of the anhydride bond. Other quantities that affect the degradation kinetics include the polymer degradation product solubility as a function of pH, and its average p$K_a$. Specifically, the degradation process by hydrolysis leading to the cross-linked polymer network breakup was carried out at 37°C in phosphate buffered saline solution at pH 7.4. An important new finding was the observation of a substantial delay period (induction), about 10 h, when water intake occurs, before the onset of the erosion. Earlier degradation experiments[4,9,13,22,26-29] were done for linear polymers only, and their phenomenological modeling was attempted.[12] These studies typically reported no indications of such a long induction period, if any, preceding a noticeable mass loss.

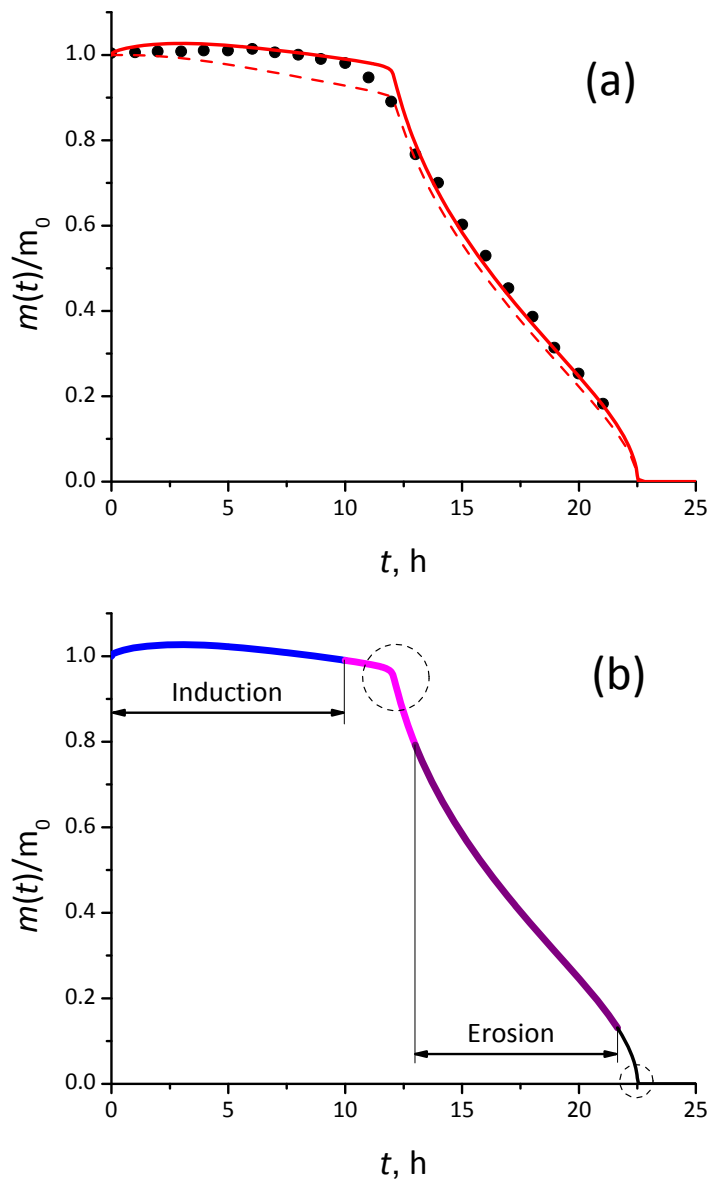

**Figure 1.** Fractional mass remaining in the undissolved sample as a function of time, including increase due to water intake, but decrease due to loss of polymer matter by erosion. (a) Black dots are the experimental data.[17] The solid line is the model fit for the total mass, whereas the dashed line is the polymer-only mass in the surviving sample. (b) Features of the model-predicted erosion profile time dependence, discussed in the text.

Here, degradation in the polymer connectivity occurs as water is taken up by the sample and the anhydride functional groups are hydrolyzed.[17] This leads to erosion once low molecular weight degradation products and possibly short oligomers are lost from a bulk polymer into the solution. We aim to explain the nature of the experimentally observed[17] delay (induction) interval preceding measurable erosion (mass loss) of the sample. Figure 1(a) illustrates this effect by showing experimental data[17] and also our model fit curves. The details of the system and the description of the modeling approach are presented in the following sections.

## 2. THEORETICAL

Earlier modeling work[6-16,30] on bulk and surface eroding polymers does not provide tools for understanding the long water-intake interval for the recently developed cross-linked amorphous polyanhydrides.[17,18,31,32] Figure 1 illustrates results of our successful modeling approach, reported here. These results also pose new interesting theoretical and experimental challenges. The goal of modeling is to understand the mechanisms and parameters of the relevant processes, allowing us to fit the observed data, e.g., Fig. 1(a); classify the various regimes of erosion, Fig. 1(b); offer predictive capabilities for novel water-degradable biomaterials design, and, importantly, suggest physical insights into the required future experimental studies.

In the model reported here, the observed extended delay time of water intake without substantial mass loss is attributed to an interesting kinetic effect of the breakdown of the rate-equation description for water reactivity. It should be noted that much of the previous modeling and experimental work examined semi-crystalline polyanhydrides. The presence of two phases adds further complexity to the erosion, since the amorphous and crystalline phases are expected to exhibit different degradation and erosion rates.[8] In contrast, the materials examined here are simpler morphologically since they are amorphous. This will simplify our approach to understanding erosion behavior. Cross-linking is another major difference here as compared to previous efforts, which were predominantly focused on linear polymers, and thus molecular weight was a dominating factor in the erosion process.

Many parameters can control the degradation and erosion kinetics. Rate constants are needed to describe the degradation of the polyanhydride bonds by hydrolysis. Diffusion constants should be considered for water and the detached degradation product(s) in the cross-linked matrix, and also for the degradation product(s) outside the sample. These parameters might be pH-dependent leading to variations in the rate of hydrolysis,[9,11,26,28,33,34] and in the solubility of the degradation products.[10,13,27] Therefore, diffusion of buffer species may also need to be considered. As the matrix degrades with time and mass loss occurs, the receding sample boundary must be defined, and the variation of concentrations across it considered. This introduces additional parameters. The observed swelling of the outer layer of the sample[17] involves yet another set of parameters for modeling.

The full details of the model will be described later. In this section we offer an outline to illustrate the findings. Let us first consider the standard rate equations, used in earlier modeling,[6,7,12,30,35] for the degradation of the network inside the sample, here a narrow slab as in the experiment,[17] with $x$ measured from its middle (and $t$ representing time),

$$\frac{\partial u_4(x,t)}{\partial t} = -k_4 u_4 u_w;$$

$$\frac{\partial u_i(x,t)}{\partial t} = k_{i+1} u_{i+1} u_w - k_i u_i u_w, \quad i = 1,2,3;$$

(1)

$$\frac{\partial u_0(x,t)}{\partial t} = k_1 u_1 u_w + D_0 \frac{\partial^2 u_0}{\partial x^2};$$

$$\frac{\partial u_w(x,t)}{\partial t} = D_w \frac{\partial^2 u_w}{\partial x^2}.$$

Here we already see a large number of parameters involved. These include the rate constants $k_{i=1,2,3,4}$ and molar concentrations $u_{i=0,1,2,3,4}$ of the 4-, 3-, 2-, 1-(cross-)linked (to the network) fragments, as well as 0-linked detached small fragments. The latter fragments are small enough to diffuse. Most, though not all of them, will be 0-linked units, diffusing with the average diffusion constant $D_0$. The reaction terms in the rate equations describe the hydrolysis that breaks up the network due to the presence of water (molar concentration $u_w$) that diffuses in the cross-linked matrix with diffusion constant $D_w$.

The outer boundary of the sample was defined at $x = X_B(t)$, such that the total amount of the cross-linked material at this *x* value dropped to some reference fraction, another parameter, *g*, here taken 7.6 % for illustration (see the next section), of the maximum for the fully 4-linked initial system at time *t* = 0,

$$\sum_{i=1}^{4} u_i(X_B(t), t) = g u_4(0). \tag{2}$$

For $x > X_B(t)$, additional modifications of the kinetic description are required, and some quantities undergo qualitative changes in properties at the boundary. We note, however, that already the set of the reaction-diffusion equations, Eqs. (1), involves 7 parameters. Some are actually known, at least approximately, specifically, $D_w$ in such a polymeric material environment.[6,36-39] Other parameters cannot be accurately determined because of the noise in the fitted data, or the model not fitting the data. This holds, for instance, for $D_0$ in the model of Eqs. (1), but not in the improved model described below. Still, we are left with several rate constants, etc., including some of the parameters defined at the boundary and outside the sample (described in the next section), and even more quantities that can be fitted in relation to the observed swelling, which we also modeled, see the next section, where all the relevant parameter values are given. Note that the data, Fig. 1(a), actually offer two well-defined time scales as the key measurable properties to fit: the delay time of the induction region, and the erosion time, i.e., the duration of the approximately linear (for our slab-shaped sample) decay once the erosion sets in, as summarized in Fig. 1(b).

A surprising finding has been that the standard rate-equation model that includes Eqs. (1), with all the added elaborations such as the behavior at the boundary and even the inclusion of swelling, etc., cannot reproduce the two experimentally observed time scales, despite a large number of adjustable parameters. As illustrated in Fig. 2 for varying values of some of the rate constants (among $k_{i=1,2,3,4}$), there is no well-defined delay time of the induction region. Adjustment of the time scale of the decay region is possible, by parameter fitting, but no comparable-duration delay region is seen. There is also a "bottleneck" effect, exemplified in Fig. 2, that only a single parameter, typically one of the rate constants, is the primary rate-limiting one, so there is generally much less freedom in the parameter determination by data fitting than one would expect from the large number of the possible model parameters. The latter property is actually typical for such models of chemical kinetics.

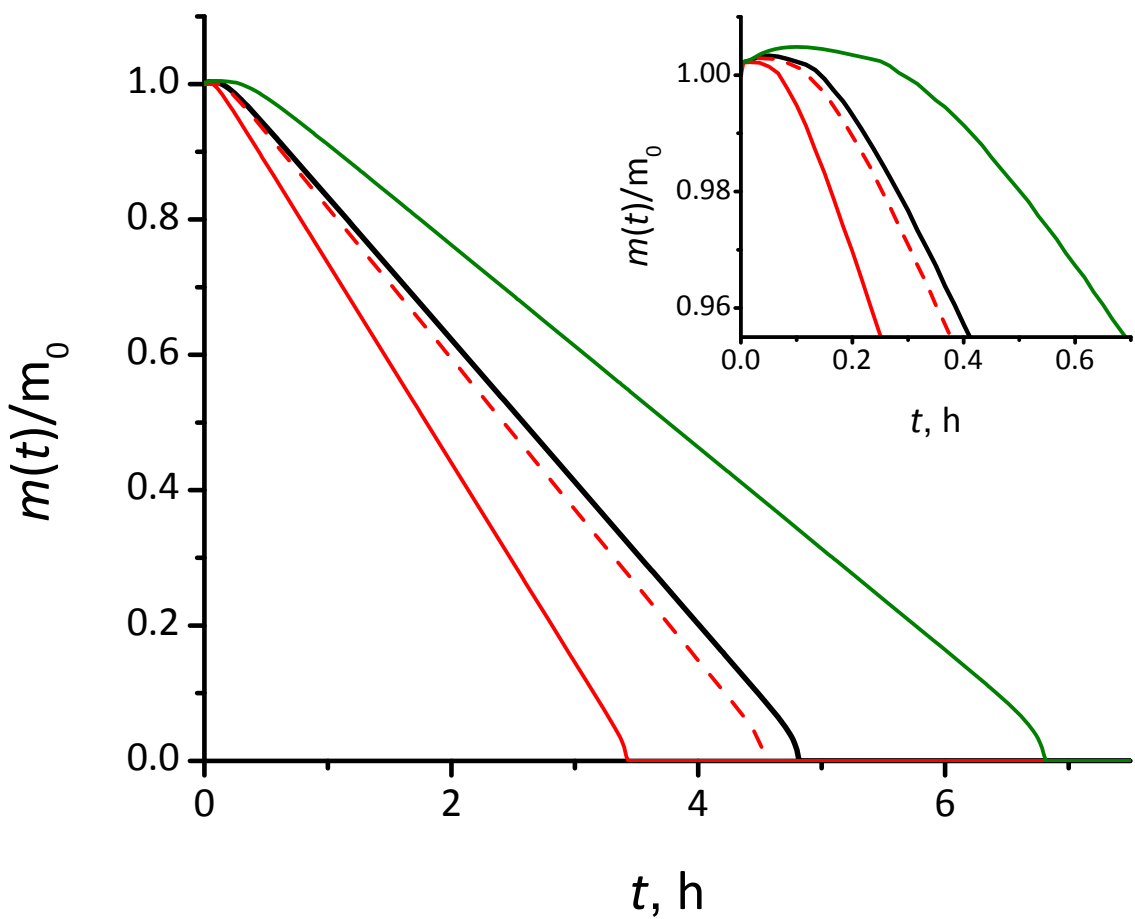

**Figure 2.** Fractional mass of the undissolved sample as a function of time in the standard-rate-equation model. The solid red line has double all the rate constants, $k_i$, as compared to the black line. The green line was obtained by instead halving all the rate constants at once. The dashed red line represents the case when only $k_2$ is doubled. The latter illustrates the bottleneck effect, because $k_2$ is not the rate-limiting quantity for the particular set of the parameter values used. The parameter value choices are explained in the text. The inset shows the details of the behavior for shorter times.

The net result is that the rate equation model fails to explain the large induction time. We note that several other experiments, which involved networks that were less cross-linked did not report the long induction time,[4,9,13,22,26,27,29,34] although some showed a delay time.[7,9,17,18,40-42] In numerous variants of the model, the rate constants in Eqs. (1) were found to be the primary rate-controlling parameters, and at the same time these equations were the only "constant feature" of the otherwise rather sophisticated model variations tried. Therefore, we are led to the conclusion that their validity should be questioned. We found that, the concentration-of-water dependence is the culprit. The model, and even its simplified variants can fit the data provided each of the reaction terms in Eqs. (1) is replaced according to

$$k_i u_i u_w \Longrightarrow k_i u_i u_w f(u_w), \quad i = 1,2,3,4, \tag{3}$$

with a single-parameter function, here taken as

$$f(u_w) = \frac{u_\infty}{u_w + u_\infty}, \tag{4}$$

phenomenologically describing the deviation of the original $u_w$-dependence of the reaction rates from linear, to saturate as a finite value, $u_\infty$, for large $u_w (\gg u_\infty)$.

We emphasize that the form used in Eq. (4) is entirely phenomenological. We took the simplest possible rational function that depends on the ratio $u_w/u_\infty$ in such a way that the linear dependence on $u_w$ in Eqs. (3), when replaced with $u_w f(u_w)$, remains linear in $u_w$ for $u_w \ll u_\infty$, but levels out as $u_w$ increases towards $u_\infty$, and saturates at $u_\infty$ for $u_w \gg u_\infty$. Thus, $f(u_w)$ represents the fraction of water that is "reactive," but its precise form can only be determined from future experiments that will probe microscopic morphological properties rather than presently studied macroscopic erosion profiles. We will further discuss this matter and comment on the implied physical insights in the concluding section.

This model modification, further discussed in the next section, yielded the curves in Fig. 1(a). The single parameter, $u_\infty$, controls the induction time. Furthermore, with the well-defined induction region reproduced, we also noted that the shape and sharpness of the transition region from induction to decay, circled in Fig. 1(b), are largely controlled by the assumed initial degree of cross-linking at the boundary and the 0-linked degraded unit diffusion constant, $D_0$, though the data are too noisy for the precise determination of the latter. The model predicts the sharp drop-off region at the end of the erosion time, which is also circled. This is difficult to verify experimentally because the sample then literally breaks up and its mass cannot be measured.

## 3. DISCUSSION

In this section we discuss model details, with emphasis on the various parameter values involved, many of which are adjustable. We then address, in the next section the proposed modification, Eqs. (3)-(4), of the standard rate-equation description, Eqs. (1), and offer a concluding discussion.

### 3.1. Parameter Values Specific to the Experiment, and Experimental Details

The polyanhydrides were synthesized[17] through the thiol-ene polymerization of pentaerythritol tetrakis(3-mercaptopropionate) (PETMP) and 4-pentenoic anhydride (PNA); see Scheme 1. Radical-mediated thiol-ene reactions exhibit 'click' chemistry characteristics, meaning that they are highly efficient, easy to perform and tolerant to the presence of other chemical functionalities.[43-45] For the reactions presented here, the PETMP (tetra-thiol monomer, 0.7 mmol) and the PNA (diene monomer, 1.4 mmol), yield a highly cross-linked network structure, as shown in Scheme 1. Since the PNA monomer contains the anhydride functionality that readily undergoes hydrolysis, these cross-linked polymers degrade in aqueous environments. The polymers were synthesized by combining the PETMP, PNA, and photoinitiator (1-hydroxycyclohexyl phenyl ketone, 0.6 mg) in a polydimethylsiloxane mold (10 × 10 × 2 mm) and irradiating with UV light (~70 mW/cm$^2$) for 15 minutes. The samples were subsequently degraded in 100 mL of phosphate buffered saline (PBS) solution at 37° C and pH 7.4. Every hour the polymer was removed from the buffer, quickly dried, weighed, and placed into fresh PBS solution.

The PETMP monomer has molar weight 488.66 g/mol, and PNA, 182.22 g/mol.[46,47] Total molar weight of the repeat unit in the cross-linked polymer is $W_p$ = 853.1 g/mol, therefore the molarity, $M_p$, is 1.31 M, where the volume density, $\rho_p$, of polymer was taken 1.12 g/ml.[17] The molar weight of pure water, $W_W = 18.0153$ g/mol, along with the density of pure water at room temperature, $\rho_w = 0.997$ g/ml, yield the molar concentration $M_w = 55.36$ M.

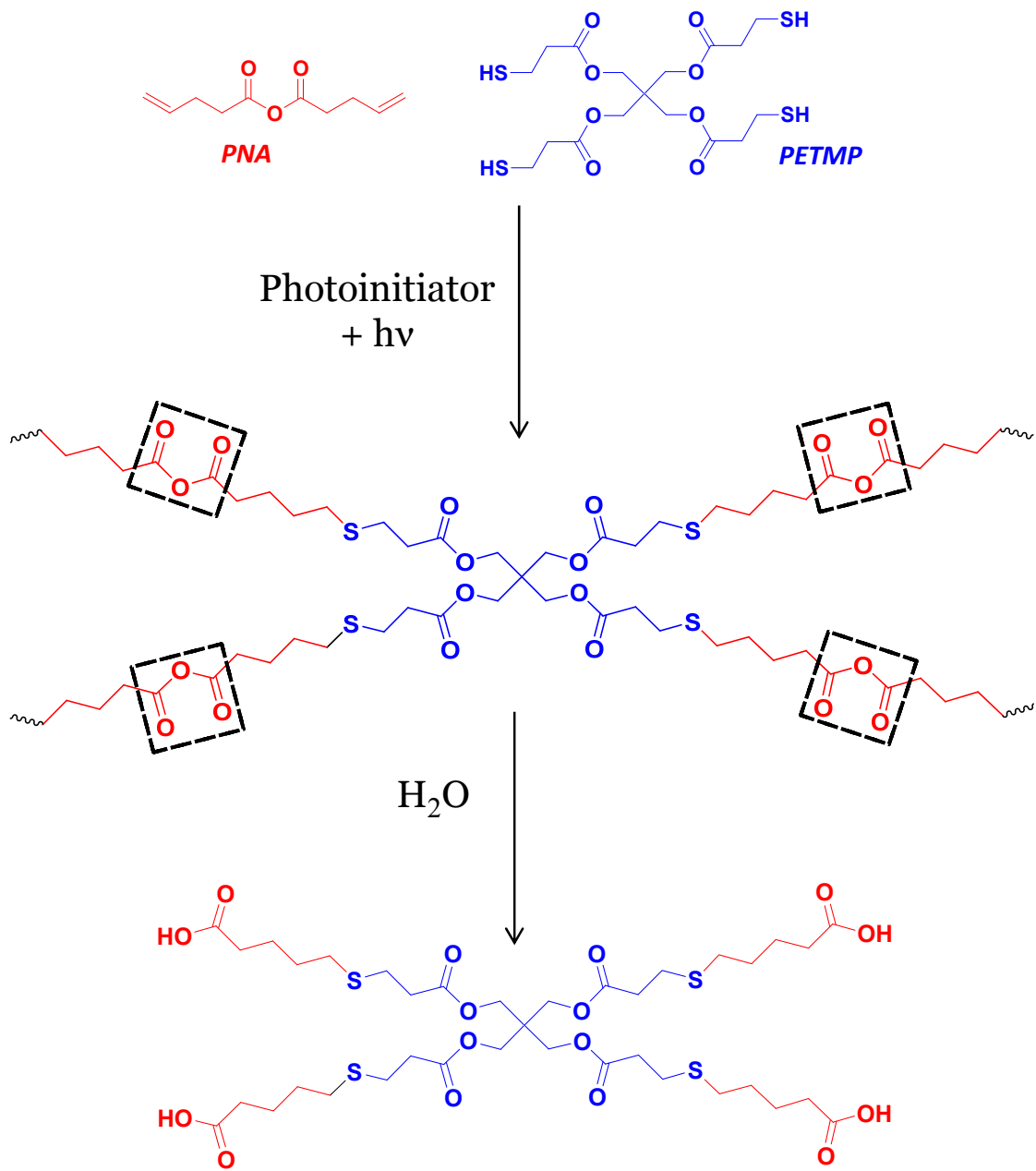

**Scheme 1.** The top images show the two types of the monomer molecules used, as described in Sec. 3.1. These are polymerized to form a cross-linked structure as illustrated in the middle image, with the dashed-line boxes highlighting the anhydride bonds. These bonds are degraded by hydrolysis to break the polymer up into small units, such as the one shown in the bottom image.

### 3.2. Definition of the Boundary

For the purpose of our model a slab (or cuboid) is chosen in order to be consistent with the experiment.[17] The $x$ coordinate is measured from its center, and the instantaneous half-thickness of the eroding slab is $X_B = X_B(t)$, starting from the initial, $X_B(0) = 1$ mm. In the $y$- and $z$-directions, the slab is much bigger than in $x$ direction (10 mm in the experiment) and therefore all the densities and parameter values, e.g., molar concentrations, reaction rates and diffusion coefficients, are assumed to be only $x$-dependent, uniform in $y$ and $z$. In the experiment,[17] the proportions of the sample remained largely unchanged and its shape was not distorted during the erosion process.

Eq. (2) shows how the boundary is defined, where the value of the parameter $g = 7.6\%$ is chosen to yield $g \cdot M_p = 100$ mM. This is a somewhat arbitrary choice that assumes that the material is entirely broken up (is no longer solid) at this relatively low density, but we found that the precise choice of $g$ has no significant effect of the erosion properties. The detached matter is not counted in the mass of the remaining sample. It ultimately degrades to small molecules by hydrolysis in the solution outside the solid sample.

### 3.3. Variation of Densities, Diffusion Properties, and Initial Conditions at the Boundary

Initially, we took almost the entire polymer matrix as 4-connected, with a small quantity of 3-connected units, with the molar concentrations,

$$u_{1,2}(x, 0) = 0,$$

$$u_3(x, 0) = M_p e^{[x-X_B(0)]/\delta} \Theta(X_B(0) - x), \tag{5}$$

$$u_4(x, 0) = M_p - u_3(x, 0),$$

where $\Theta$ is the step function that vanishes past $X_B(0)$. The parameter $\delta$ is small; we used $10^{-5}$ m in Fig. 1(a). It was found to control the rounding of the induction to erosion transition region: A sharper upper-circled corner in Fig. 1(b) is obtained with a smaller $\delta$. This is demonstrated in Fig. 3, which also shows the corresponding initial conditions. Physically the value of $\delta$ is likely set as the sample slightly degrades due to the air moisture after it was prepared, but before immersing it into the aqueous buffered solution. We also note that initially, water is not present within the sample,

$$u_w(x, 0) = M_w \Theta(x - X_B(0)). \tag{6}$$

During the degradation process the bonds are hydrolyzed (broken). Small enough detached units diffuse through the structure.

We consider only one half of the slab, at $0 < x < X_B(t)$, and therefore reflecting boundary condition was used at $x = 0$,

$$\left.\frac{\partial u_w(x,t)}{\partial x}\right|_{x=0} = \left.\frac{\partial u_0(x,t)}{\partial x}\right|_{x=0} = 0. \tag{7}$$

The following boundary conditions are assumed at $x = X_B(t)$, to have a source of water diffusing into the polymer matrix and a sink for the detached monomers,

$$u_w(x,t)|_{x=X_B(t)} = pM_w,$$

$$u_0(x,t)|_{x=X_B(t)} = 0, \tag{8}$$

where $p$ is the fraction of the amount of water that can be taken up by the polymer matrix due to its limited porosity, i.e., this parameter limits the maximum molar concentration of water inside the polymer matrix. This value was fitted to get the increase in the mass of the sample in the induction region, $p = 0.05$, though its determination is not precise because the data[17] are noisy.

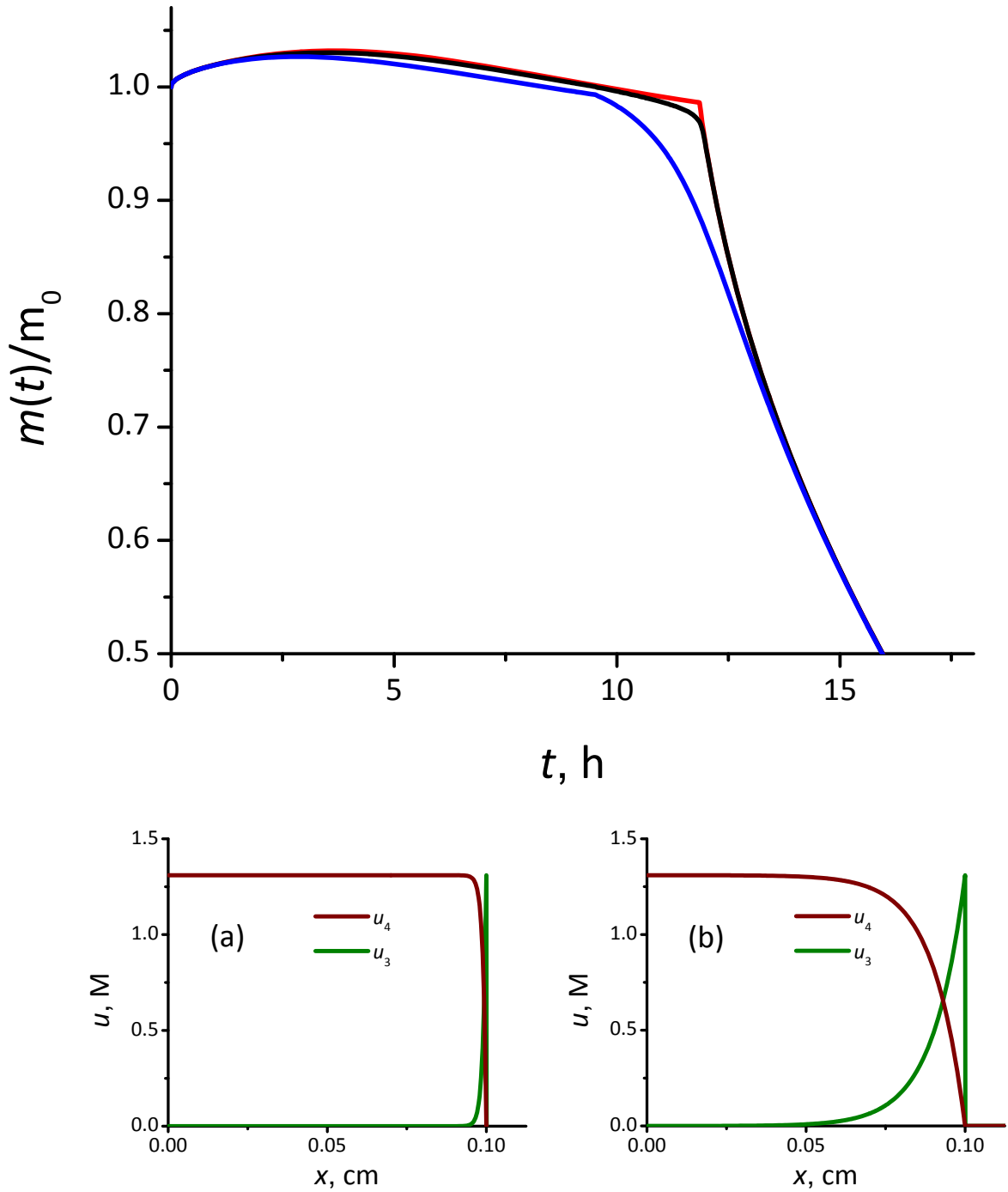

**Figure 3.** Effect of the initial small fraction of the 3- vs. 4-crosslinked units at the sample boundary, see Eqs. (5), on the rounding of the induction to erosion transition. Top panel: the red line is for $\delta = 0$; the black line corresponds to initial conditions (a) $\delta = 10^{-5}$ m; the blue line is for (b) $\delta = 10^{-4}$ m. The initial conditions for the 3- vs. 4-crosslinked unit concentrations are shown in the two bottom panels, (a) and (b), respectively.

### 3.4. Comments on Swelling

As mentioned in the preceding section, before accepting the surprising result that the rate equations involving water concentration are not applicable without the reactivity suppression factor, Eqs. (3) and (4), we tried other modifications of the model that introduced additional adjustable parameters. Specifically, we considered swelling of the outer sample layer, observed during the experiment.[17] The swelling was assumed to occur starting at $X_S(t) < X_B(t)$. We assumed that swelling is significant once the network is mostly 1- and 2-connected due to bond breakage, i.e., $X_S(t)$ is defined as the smallest $x$-value beyond which $u_1 + u_2$ exceeds a certain threshold. Without going into details, we note that when the decompression ratio is considered as another adjustable parameter, varied from 1 (no swelling) up to 1.5, no significant effect on the kinetics is observed, and specifically, no large induction period can be obtained in the original rate-equation description.

### 3.5. Fitting the Parameter Values, and the Bottleneck Effect

The diffusion coefficient of water in the polymer solid structure was taken $D_w = 10^{-8} \text{cm}^2/\text{s}$.[6,36-39] The diffusion coefficient of the detached 0-connected units was considered an adjustable parameter, with the best fit value $D_0 = 0.3 \cdot 10^{-8} \text{cm}^2/\text{s}$.

As discussed in the preceding section, there is a "bottleneck effect" in how various rate constants affect the polymer degradation process. The smallest of all $k_i$ effectively limits the degradation rate, i.e., the overall rate of conversion of 4-connected units to 0-connected. In fact, we found that to fit the data shown in Fig. 1(a), to a good approximation a single value can be used for all the rate constants because of this effect. The best fit obtained for the successful model, with the modification shown in Eqs. (3) and (4), was

$$k_{i=1,2,3,4} = 4.56 \cdot 10^{-3} \text{ s}^{-1}\text{M}^{-1}. \tag{9}$$

Finally, the parameter $u_\infty$, see Eqs. (3) and (4), had the fitted value

$$u_\infty = 37 \text{ mM}. \tag{10}$$

The curves demonstrating of the "bottleneck effect" in Fig. 2 were calculated with all the same parameters, but without the reduction in the water reactivity (i.e., effectively with $u_\infty = \infty$).

**3.6. Reaction Profile Snapshots and Computation of the Sample Mass**

To demonstrate how the degradation process occurs in our model, in Fig. 4 we show the concentrations snapshots (calculated with the fitted parameter values) at different times, $t$, indicated above each plot. These include the snapshot at $t = 0$ or initial conditions, the same as shown on Fig. 3(a). Such time-dependent concentrations were used to compute the fractional mass remaining in the undissolved sample as a function of time (see Fig. 1), according to

$$m(t) = 2\mathrm{A}\int_0^{X_\mathrm{B}(t)}\{\mathrm{W}_\mathrm{p}[\textstyle\sum_{i=0}^{4} u_i(x,t)] + \mathrm{W}_\mathrm{w} u_\mathrm{w}(x,t)\}\,dx, \qquad (11)$$

where A is a slab cross-section area. At time $t = 0$, the mass equals

$$m_0 = 2\mathrm{A}\mathrm{W}_\mathrm{p} X_\mathrm{B}(0). \qquad (12)$$

**3.7. Comments on the Sharp Drop-off Region**

In the present model, the function $X_\mathrm{B}(t)$ actually reaches zero at a finite time, as seen in a circled region in Fig. 1(b). This simply reflects the fact that the model ignores any density fluctuations and non-uniformities along the $y$ and $z$ directions. In reality, once the sample becomes sufficiently thin such fluctuations will round up the short drop-off to zero in the mass erosion profile, but our model does not account for this.

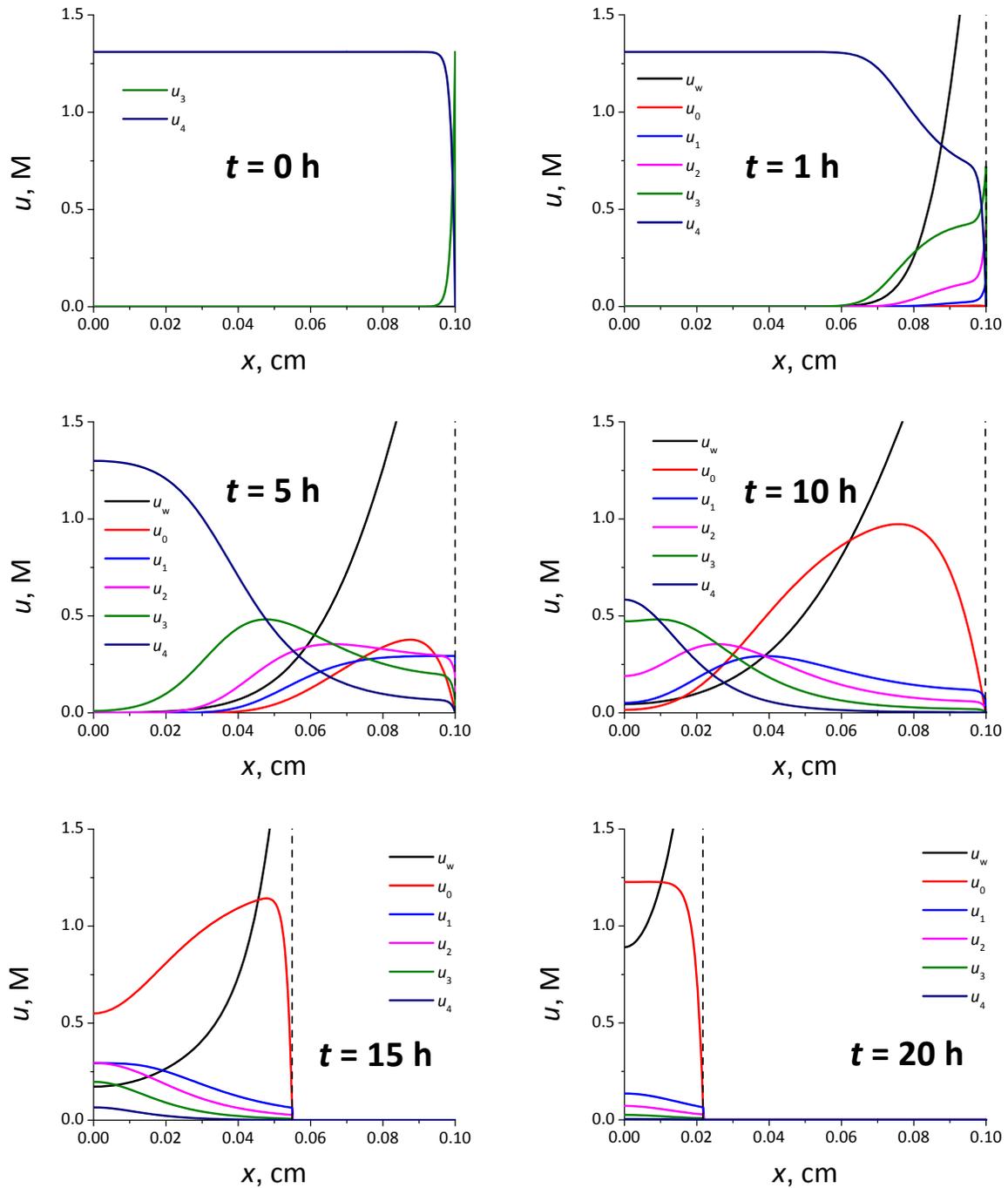

**Figure 4.** Spatial dependence of the concentrations of units of various degree of crosslinking, as well as of water at different times up to $t = 20$ h. The vertical dashed lines mark the sample boundary at $X_B(t)$. Note that initially we assume a small fraction of 3-cross-linked units in addition to mostly 4-cross-linked units, according to Eq. (5). Even a rather small amount of water present in the material suffices to significantly increase their relative density, as can be seen for $x \lesssim 0.02$ cm at time $t = 5$ h.

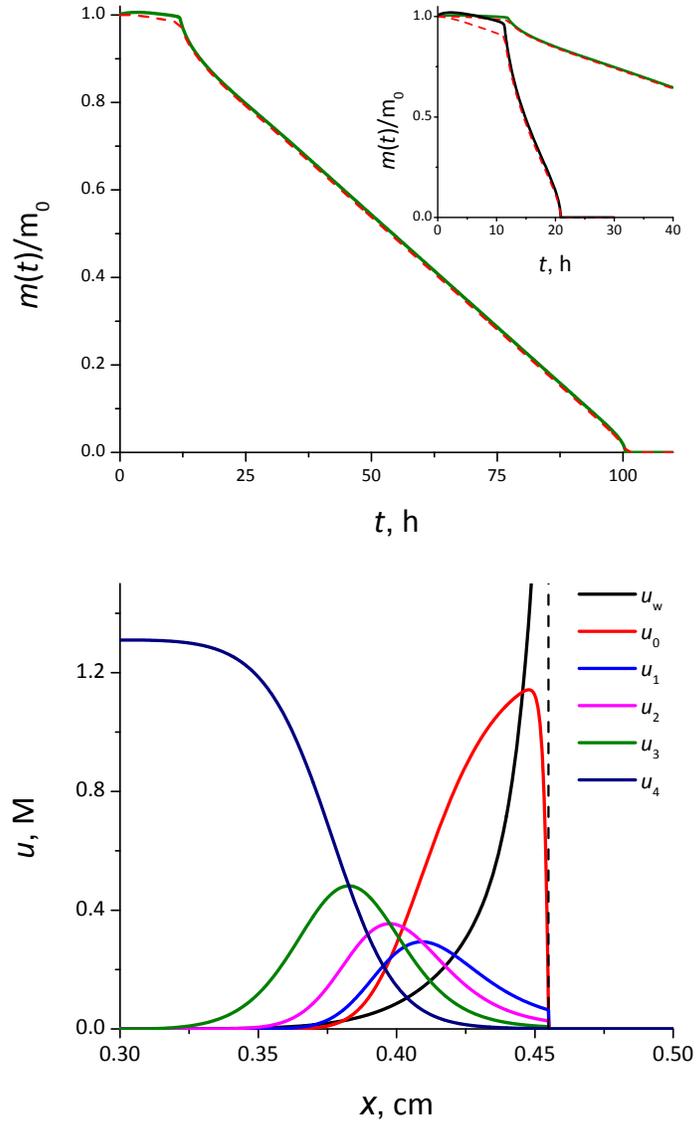

**Figure 5.** Top panel: Interestingly, the model predicts that the induction time is not significantly dependent on the sample thickness, as expected for surface erosion, whereas the erosion, once it sets in, takes longer time for thicker samples. Here this is illustrated for the initial sample size of 1 cm. The solid green line gives the model-calculated fraction of the remaining mass for such a sample. The inset compares this to the same result from Fig. 1(a): the solid black line (sample size 2 mm). The dashed red lines represent the mass of the polymer matrix with water mass subtracted. Bottom panel: The concentrations $u_i$ as a snapshot at $t = 15$ h, approximately once the constant-rate erosion sets in. The dashed black vertical line gives the sample's boundary position, $X_B$, at that time.

## 3.8. Illustrative Calculation for a Different Polymer Sample Size

A sample with different geometry (or size) will have a different overall erosion rate, for example, see Fig. 5, which should be compared to Fig. 1(a). Here we chose a cuboid with the initial thickness of 1 cm, but still with the assumption that the sample size in the directions $y$ and $z$ is much larger than in $x$ direction. All the other parameters remain the same.

## 4. CONCLUSION

The following two observations highlighted in Fig. 5, confirm that our model predicts surface rather than bulk erosion for the considered system. First, the induction interval remains approximately the same, and, second, the erosion, once it sets in, results in an approximately linear with time loss of mass (a geometrical property for a surface-eroding thin-slab shape).

The key finding in the present study has been that, despite several adjustable parameters available for data fitting, the observed long induction interval can only be explained by allowing for the reduction in water reactivity as its concentration locally increases inside the sample. Here this was described by assuming a phenomenological rational-function expression, see Eqs. (3)-(4), with a single adjustable parameter, $u_\infty$. All the rates in the standard rate equations, Eqs. (1), that were proportional to the water concentration, $u_\text{w}$, are made instead proportional to

$$u_\text{w} f(u_\text{w}) = \frac{u_\text{w} u_\infty}{u_\text{w} + u_\infty}. \tag{13}$$

We again emphasize that the choice of the specific functional form for $f(u_\text{w})$ in Eqs. (3), (4), (13), was made out of convenience. In order to fully model this phenomenologically discovered reduction in the water reactivity, we need not just modeling but new microscopic experimental data, presently not available for these systems. The following discussion offers some physical insights into possible mechanisms for such behavior.

Indeed, our findings suggest interesting avenues of research. Once evidence was found that the induction region correlates with the water reactivity in hydrolysis in the present system not being entirely "diffusion-controlled," we can seek physical reasons for this anomalous behavior. One explanation could be that the network is particularly dense, thus preventing fast enough local water equilibration by diffusion once its concentration increases for the reaction rates to pick up. However, statistical-mechanics considerations[48] make this explanation unlikely, unless water diffusion in this network is itself anomalous. Another explanation could be that a dense network prevents resupply of buffer by diffusion from outside the sample, and the reaction slows down due to local uncompensated pH changes. A more likely explanation of the observed limitation in water reactivity in a way points to an opposite effect: As the network is eroded by hydrolysis, and more water is taken up, some of it will be in large enough defects/crevices that only a fraction of the water will be surface-reacting with the surrounding network. Some evidence for such surface-only reactivity has been noted in a different context.[10,13] All these options suggest the need for detailed microscopic process models to be guided by future structural morphological, and perhaps mechanistic experimental studies supplementing macroscopic erosion data, which are presently not yet available.

## ACKNOWLEDGMENTS

We acknowledge funding by NSF under award CBET-1403208. We also thank Garrett Liddil, Halimatu S. Mohammed, Damien S. K. Samways, and Brittany L. Snyder, for their collaboration.